\documentclass[conference]{IEEEtran}
\IEEEoverridecommandlockouts
\usepackage{cite}
\usepackage{tabularx}
\usepackage{amsmath,amssymb,amsfonts}
\usepackage{algorithmic}
\usepackage{graphicx}
\usepackage{textcomp}
\usepackage{xcolor}
\usepackage{enumitem}
\usepackage{tabularx} 
\usepackage{longtable}
\usepackage{xspace}
\usepackage{longtable}
\usepackage{lipsum}  
\usepackage{booktabs} 
\usepackage{afterpage} 
\usepackage{tikz}
\usetikzlibrary {positioning}
\usetikzlibrary{shapes.geometric, arrows}
\usepackage{multirow}
\usepackage{float} 

\usepackage{url}
\def\BibTeX{{\rm B\kern-.05em{\sc i\kern-.025em b}\kern-.08em
    T\kern-.1667em\lower.7ex\hbox{E}\kern-.125emX}}

\newcommand{\ie}{\textit{i.e.,}\xspace}
\newcommand{\eg}{\textit{e.g.,}\xspace}

\newcommand{\etal}{\textit{et al.}\xspace}

\tikzstyle{block} = [rectangle, rounded corners, minimum width=3cm, minimum height=1cm, text centered, draw=black, fill=blue!10]
\tikzstyle{arrow} = [thick,->,>=stealth]
\begin{document}

\title{Machine Learning Solutions Integrated in an IoT Healthcare Platform for Heart Failure Risk Stratification}

\author{\IEEEauthorblockN{
Aiman Faiz$^+$, Claudio Pascarelli$^\#$,  Gianvito Mitrano$^\#$, \\ 
 Gianluca Fimiani$^+$, Marina Garofano$^+$, Mariangela Lazoi$^\#$,  \\ Claudio Passino$^\otimes$, Alessia Bramanti$^+$
}
\IEEEauthorblockA{
\textit{$^\#$University of Salento}, \textit{$^\otimes$Scuola Superiore Sant’Anna di Pisa} , \textit{$^+$University of Salerno}}
}

\maketitle

\begin{abstract}
This paper presents a predictive model founded on Machine Learning (ML) techniques to identify patients at Heart Failure (HF) risk. This model is an ensemble learning approach, a modified stacking technique, that uses three ML models leveraging clinical and echocardiographic features and then a meta-model to combine the predictions of these three models. We assessed the model on a real dataset, and our results suggest that it performs well in the stratification of patients at HF risk. Specifically, we obtained high sensitivity (91\%) and good accuracy (78\%). 
To better study the value and potentiality of our model, we also contrasted its results with those obtained by using several baseline ML models. The preliminary results indicate that our predictive model outperforms these baselines that flatly consider features, \ie not grouping features in clinical and echocardiographic features. The proposed model is a part of the PrediHealth research project. This project integrates telemedicine, mobile health solutions, and predictive analytics into an IoT healthcare platform to manage patients at HF risk.

\end{abstract}


\section{Introduction}

In modern healthcare systems, efficiently stratifying individuals\footnote{It consists in categorizing them into different risk groups based on their health status, disease progression, or other factors.} based on their risk profiles and predicted health trajectories is crucial for optimizing medical interventions, resource allocation, and personalized treatment strategies. The increasing availability of electronic health data, wearable sensor data, and advanced Machine Learning (ML) techniques has created new opportunities for predictive modeling in patient management. However, despite these advancements, significant challenges remain in ensuring the predictive models' accuracy. In particular, the heterogeneity of patients and the complexity of disease progression demand stratification approaches that can adapt to diverse healthcare environments~\cite{alzheimers2022}.


The human heart works like a pump, sending blood and oxygen around the body. A major cause of morbidity and mortality worldwide is chronic Heart Failure (HF), which is defined by the heart's inability to pump blood efficiently. Ageing of population and lifestyle risk factors such as smoking, poor diet, and high blood pressure are closely associated with the increase in the incidence of heart disease. Current techniques often fail to provide prompt and precise identification of heart disease, even with advances in medical diagnostics \cite{odrobina2024heartfailure, toumpourleka2021risk}. It is also worth mentioning that cardiovascular diseases are associated with the highest number of deaths among Non-Communicable Diseases, amounting to 17.9 million deaths annually, followed by cancers (9.3 million), chronic respiratory diseases (4.1 million), and diabetes (2 million, including kidney disease linked to diabetes)~\cite{BramantiEtAllHealthCare2025}.

The PrediHealth project aims to address the challenges above by integrating telemedicine, mobile health (m-health) solutions, and predictive analytics into an IoT healthcare platform. Specifically, it leverages an interoperable IoT web-based application, a telemonitoring kit equipped with medical devices and environmental sensors, and ML/AI-driven predictive models to support clinical decisions~\cite{cassieri2025predihealthtelemedicinepredictivealgorithms}. As it is easy to follow, at the core of the PrediHealth project lies the predictive model--designed for the stratification of patients at HF risk--that is presented in this paper for the first time. This model is based on an ensemble learning approach--a modified stacking technique--that uses three ML models leveraging clinical and echocardiographic features and then a meta-model to combine the predictions of these three bottom models. In this paper, we detail our model and its initial empirical assessment, as well as the possible implications for our results inside and outside the goal of the PrediHealth project.  We assessed the validity of the model by conducting an initial empirical study. The obtained results suggest that the model performs pretty well. We obtained high sensitivity (91\%), ensuring that such a model nearly identifies all HF patients. As for accuracy, we obtained 78\%. It is acceptable in the PrediHealth project, given the priority of identifying patients at risk of HF to be then included in a telemonitoring program, to act promptly to let them receive medical attention by using the IoT healthcare platform developed in the project. We also compared our model against three baseline models (\eg Decision Tree and Support Vector Regression). Our initial outcomes suggest that the use of our model outperforms the baseline models (\eg in terms of the accuracy and the sensitivity). 

\textbf{Paper Structure.} 
In Section~\ref{sec:StateOfArt}, we introduce the PrediHealth project and present related research and background information, while the design of the predictive model is shown in Section~\ref{sec:model}. The design of the study to empirically assess our predictive model and the results of this assessment are reported in Section~\ref{sec:design} and Section~\ref{sec:results}, respectively. In this latter section, we also discuss the results and delineate possible implications for our research. We conclude the paper in Section~\ref{sec:conclusion} with final remarks and future work. 

\section{Background and Related Work} \label{sec:StateOfArt}

In this section, we first provide information on the PrediHealth project. It is the project for which the predictive model presented in this paper has been developed. Then we continue motivating our work and discussing related research. We conclude with a brief introduction to the ML models used in our work.

\subsection{The PrediHealth Project}

The high-level workflow of PrediHealth begins with the stratification of individuals at risk of HF using AI/ML-powered predictive models. Individuals identified as patients at HF risk are then equipped with wearable and environmental sensors, which are seamlessly integrated into a telemonitoring system based on an IoT-enabled web platform. Accordingly, the primary objectives of the projects are:

\begin{description}
    \item[] \textbf{O1: Develop an interoperable web-based IoT platform.} The platform has to establish a secure, scalable, and standardized digital infrastructure for the seamless collection, transmission, and analysis of health data from telemonitoring devices. This platform facilitates real-time monitoring of HF patients by integrating wearable medical devices and environmental sensors, ensuring continuous and remote patient monitoring.
    
    \item[] \textbf{O2: Design a telemonitoring kit.} This objective focuses on the development of a telemonitoring kit specifically designed for patients with HF, integrating advanced (wearable and not) medical devices and environmental sensors to enable continuous health monitoring.  
    
    \item[] \textbf{O3: Develop a multimarker score to support decision-making.} The goal is twofold: (i)~building AI/ML-powered models to risk stratification of patients vulnerable to acute exacerbations or disease progression for HF and (ii)~defining thresholds and scores for risk factors associated with HF exacerbations of patients when using telemonitoring kits (\ie medical devices and environmental sensors identified in O2). 
\end{description}

We restricted ourselves to a few aspects of the project for space reasons and scant relevance to the overarching goals of the research presented in the paper. Readers interested in further details can refer to \cite{cassieri2025predihealthtelemedicinepredictivealgorithms}. 

\subsection{Predictive Models in Digital Health}
Predictive models play a crucial role in the management of chronic diseases such as HF by enabling early detection, risk stratification, and personalized treatment for patients. These models analyze large datasets to support proactive interventions, reduce hospitalizations, and improve patient outcomes, making predictive modeling a key component of modern telemedicine and, more specifically, telemonitoring. Existing research has primarily focused on developing predictive models using electronic health records and demographic variables to enhance early detection and risk stratification. For example, Medhi \etal~\cite{medhi2024artificial} conducted a review by selecting 150 studies from databases such as PubMed, Google Scholar, and the Cochrane Library. Their synthesis detailed how AI/ML-powered models—spanning classical methods like SVMs, Random Forest, and decision trees to advanced deep learning approaches—can process large datasets from wearables, imaging, and other clinical sources to enable early detection, risk stratification, and personalized treatment of heart failure. Their work highlights the potential of these predictive models to reduce hospitalizations and improve patient outcomes and emphasizes the transformative role of AI in modern telemedicine and telemonitoring. 

Traditional models, such as trajectory-based disease progression models and multivariable Cox regression, have been used to identify key predictors of mortality and morbidity, including age, diabetes, and left ventricular ejection fraction~\cite{odrobina2024heartfailure}. These models often emphasize heart failure with reduced ejection fraction (HFrEF), leaving a gap in predictive capabilities for preserved ejection fraction (HFpEF) \cite{toumpourleka2021risk}.

Other researchers have explored ML-driven models for HF prognosis, utilizing algorithms like logistic regression, decision trees, random forests, support vector machines (SVMs), and ensemble classifiers. For instance, Pinem \etal~\cite{Pinem2024IntegratingMM} demonstrated that a Voting Classifier combining Logistic Regression and SVM achieved the highest accuracy (88.04\%) and ROC\_AUC score (88\%), with blood pressure and cholesterol emerging as key predictive factors. Similarly, Guo \etal~\cite{guo2020heart} employed ML techniques to predict HF outcomes using EHR data, incorporating medical notes, laboratory results, and imaging data to achieve expert-level diagnostic accuracy. These studies underscore the role of ML in early HF detection, risk assessment, and personalized intervention strategies. Recently, Visco \etal~\cite{visco2024explainable} developed a predictive model for cardiovascular diagnostics. Their study identified correlations between worsening HF and three key parameters: Creatinine, systolic pulmonary artery pressure (sPAP), and coronary artery disease(CAD). To enhance interpretability, they introduced a GP-based classifier and extracted from it a formula that achieved almost 97\% accuracy, outperforming traditional ML~models. However, the authors themselves identified how the use of only four features, despite the high accuracy, could leave out other parameters that could be important in the final prediction from the doctor's final evaluation. 

Despite these advancements, existing HF risk models face notable challenges. Voors \etal~\cite{voors2017development} showed inconsistencies in multinational cohort data, which affect model generalizability, while Adler~\etal~\cite{adler2020improving} pointed out limitations in single-center studies, particularly the exclusion of elderly populations. Moreover, many ML models struggle with imbalanced datasets and lack interpretability, which hinders their adoption~\cite{Califf2013PredictiveMI}.

The research presented in this paper aims to bridge these gaps by developing an ML-based predictive model that integrates diverse patient datasets, including HFpEF cases, to enhance generalizability. The proposed model not only refines risk stratification but also enhances clinical applicability by ensuring interpretability and reliability across heterogeneous patient populations. In addition, our predictive model is integrated within advanced solutions in the context of CPS/IoT fields\footnote{While CPS encompasses a broader range of networked systems, IoT specifically focuses on internet-enabled connectivity.} by integrating state-of-the-art technologies such as telemonitoring and m-health to facilitate the integrated hospital-territory-home management of HF patients. Finally, integrating three models within a metamodel offers several advantages over a single ML model trained on all features. We detail these advantages in Section~\ref{sec:worflow}.

\subsection{Used ML-Models}

We used two machine learning models: the Logistic Regression Classifier and the Random Forest Classifier. 
The former is a linear model widely applied in binary classification problems~\cite{log_regression_history}. It applies a weighted sum of the input features and uses the logistic function to estimate the probability of a given class. Due to its simplicity, it is computationally efficient and generally less prone to overfitting, especially in low-dimensional spaces. These characteristics, along with the inherent interpretability of the model's results (i.e., how easily a person can trace and explain the output based on the input features), make the Logistic Regression Classifier particularly valuable in clinical settings~\cite{log_regression_med}, where understanding model decisions is of central importance. 
In contrast, the Random Forest Classifier is a non-linear, ensemble-based method that constructs a collection of decision trees during the training phase and aggregates their outputs to form a final prediction~\cite{rf_pres}. This approach enables the model to capture intricate patterns and interactions across features, particularly in heterogeneous or high-dimensional datasets. While Random Forests are generally robust and versatile, they tend to lack the transparency of linear models, which may be a drawback in domains where interpretability is a key requirement.

\section{Design of the Predictive Model}\label{sec:model}

In this section, we first introduce the dataset employed for the training and the testing of our model, and then we describe it in detail. 

\subsection{Dataset}
The used dataset contains real medical data. That is, it comprises clinical records of patients diagnosed with heart failure, collected as part of a structured medical registry. It encompasses a diverse set of 1040 patient records, each characterized by unique ID and 65 features that provide a comprehensive view of patient demographics, clinical history, treatment details, and outcomes. The dataset includes key variables such as age, height, weight, and body mass index (BMI), along with clinical indicators like ejection fraction (EF), New York Heart Association (NYHA) classification, and specific diagnoses, including ischemic, hypertensive, and dilated cardiomyopathy. Additionally, dataset documents prescribed medications, including ACE inhibitors, Beta-blockers, and aldosterone antagonists, offering insights into pharmacological management. Longitudinal follow-up data track patient progress, recording mortality outcomes such as death, heart failure exacerbation, and interventions like heart transplants or ventricular assist device (VAD) implantation. The dataset also contains, \eg other bio-chemical pieces of information like Na+ (Sodium) or Uric Acid.


\subsubsection{Features}
We extracted 33 out of 65 available features in the dataset, including clinical and echocardiographic parameters relevant to cardiac evaluation based on the experience of some of the authors in the HF field and our knowledge of the state of the art~(\eg \cite{visco2024explainable}). The chosen characteristics encompass demographic attributes such as age, BMI, and sex, along with essential clinical indicators such as ejection fraction (EF), NYHA classification, NT-proBNP, creatinine, glucose, and hemoglobin (Hb). Additionally, the dataset contains information on comorbidities such as hypertension, diabetes, dyslipidemia, and respiratory diseases. The selection of features was based on the experience some of the authors gained in the field of HF, taking into account the purposes of the research presented in this paper and those of the PrediHealth project (\eg stratifying individuals into patients at risk or not). Specifically, we decided to keep all data related to a standard clinical and instrumental cardiological examination (without null values). Therefore, key echocardiographic measurements were retained, including left ventricular mass index (LVMI), ventricular dimensions, atrial size, posterior wall thickness, septum thickness, right ventricular diameter, and tricuspid annular systolic excursion (TAPSE). Electrical abnormalities on electrocardiographic evaluation, such as left and right bundle branch blocks (BBSx, BBDx), atrial fibrillation (FA), atrial flutter, and pacemaker presence (PM), were also included. Height and weight were removed since they are inherently represented by the BMI feature. In contrast, other variables such as sodium, potassium, uric acid, urea, free triiodothyronine (FT3), and free thyroxine (FT4) were excluded as they were considered secondary to the main goal of our research. 
It is important to note that the \textit{Diagnosis} feature was divided into two categorical features to enhance the granularity of the information and allow for a clearer differentiation between the identified diagnoses.
Some other features were deliberately excluded due to the high percentage of missing values. Given the medical nature of the dataset, we opted not to impute missing values with synthetic or inferred data to preserve the integrity of the analysis and avoid introducing potential biases.    

Medication usage data, including beta-blockers, ACE inhibitors, and aldosterone antagonists, were incorporated to assess treatment effects. In addition, two derived characteristics were included: ``lifespan,'' representing the number of days between diagnosis and outcome, and ``label.'' The features, grouped in ``Clinical Features" and ``Echocardiographic Features," that serve as the foundation of our predictive model are detailed in Table~\ref{tab:feature_description}. 

The feature categorization into Clinical and Echocardiographic Features reflects their complementary roles in HF assessment. Clinical Features capture systemic health, metabolic status, and hemodynamic regulation, which are essential for identifying HF risk factors. Key prognostic markers like NT-proBNP, creatinine, and NYHA classification provide insight into HF progression~\cite{ponikowski2016esc}. Echocardiographic Features offer direct evaluation of cardiac structure and function. Metrics such as EF, LVMI, and TAPSE help distinguish HF phenotypes and guide treatment. The inclusion of ECG abnormalities like bundle branch blocks and atrial fibrillation enhances prognosis by detecting conduction disturbances. This grouping integrates systemic and cardiac indicators, improving risk stratification.

\begin{table*}[]
\caption{Description of the features used in the dataset.} 
    \label{tab:feature_description} 
\begin{tabular}{lll}
\hline
\textbf{Feature}       & \textbf{Description}                                                            & \textbf{Type}                      \\ \hline
\multicolumn{3}{l}{\textbf{Clinical Features}}                                                                         \\ \hline
Primary\_Diagnosis     & Primary diagnosis of the patient                                                & Categorical                        \\
Secondary\_Diagnosis   & Divided into Primary and Secondary if two diagnoses are present                 & Categorical                        \\
HFpEF                  & Heart failure with preserved ejection fraction classification                   & Categorical                        \\
EF                     & Ejection Fraction (\%) – Left ventricular pumping function                      & Numeric                            \\
NYHA                   & New York Heart Association (NYHA) class (I-IV), measures heart failure severity & Categorical                            \\
Age                    & Age of patients in years                                                        & Numeric                            \\
BMI                    & Body Mass Index                                                                 & Numeric                            \\
Sex                    & Sex of patients                                                                 & Categorical (1 = Male, 0 = Female) \\
Hypertension           & High blood pressure                                                             & Categorical (1 = Yes, 0 = No)      \\
Dyslipidemia           & Abnormal lipid levels                                                           & Categorical (1 = Yes, 0 = No)      \\
Diabetic               & Blood sugar disorder                                                            & Categorical (1 = Yes, 0 = No)      \\
Bronchopneumonia       & Lung disease                                                                    & Categorical (1 = Yes, 0 = No)      \\
Beta-Blocker           & Use of beta-blockers                                                            & Categorical (1 = Yes, 0 = No)      \\
ACE\_SART              & Use of ACE inhibitors or sartans                                                & Categorical (1 = Yes, 0 = No)      \\
Anti-Aldosterone       & Use of aldosterone antagonists                                                  & Categorical (1 = Yes, 0 = No)      \\ \hline
\multicolumn{3}{l}{\textbf{Echocardiographic Features}}                                                                                       \\ \hline
PARETE POST            & Posterior heart wall thickness                                                  & Numeric                            \\
SETTO                  & Septum thickness                                                                & Numeric                            \\
LVES\_DIAM             & Left ventricular end-systolic diameter                                          & Numeric                            \\
LVED\_DIAM             & Left ventricular end-diastolic diameter                                         & Numeric                            \\
VDx (PARAST)           & Right ventricular diameter                                                      & Numeric                            \\
LVMI                   & Left ventricular mass index                                                     & Numeric                            \\
ASx                    & Left atrium size                                                                & Numeric                            \\
TAPSE                  & Tricuspid annular systolic excursion                                            & Numeric                            \\
RS                     & Respiratory syndrome                                                            & Categorical (1 = Yes, 0 = No)      \\
BBSx                   & Left bundle branch block                                                        & Categorical (1 = Yes, 0 = No)      \\
BBDx                   & Right bundle branch block                                                       & Categorical (1 = Yes, 0 = No)      \\
NT-proBNP              & Heart failure biomarker                                                         & Numeric                            \\
Blood Creatinine Level & Kidney function marker                                                          & Numeric                            \\
Glucose                & Blood sugar level                                                               & Numeric                            \\
FA                     & Atrial fibrillation                                                             & Categorical (1 = Yes, 0 = No)      \\
Flutter                & Atrial flutter                                                                  & Categorical (1 = Yes, 0 = No)      \\
PM                     & Pacemaker presence                                                              & Categorical (1 = Yes, 0 = No)      \\
Hb                     & Hemoglobin level                                                                & Numeric                            \\ \hline
\end{tabular}
\vspace{-0.4cm}
\end{table*}

\subsubsection{Pre-processing}
Before using the dataset, we preprocessed it to ensure data consistency and reliability. Therefore, the goal of pre-processing was to include the highest number of data and have, at the same time, a balance between patients at HF risk and not.  
The following pre-processing steps were performed: 

\begin{itemize}

    \item Date values were converted into a standardized date-time format, and a new feature, ``lifespan," was introduced to represent the number of days between the initial characterization date and the recorded outcome date (``death").

    \item We hold patients for whom we had information on their death or the follow-up period was larger or equal to three years. At the end of this step, 464 patients remained in the dataset.
       

     \item Patients with at least one missing value in the features used by our model were removed from the dataset. At the end of this step, we removed 99 patients from the dataset, and then 365 patients remained.


    \item Patients with inconsistent values or placeholder values in terms of classification (intermediate values) have been removed. At the end of this step, we removed 8 patients from the dataset, and we reached the final dimension of the dataset of 357 patients


    \item As for the outcome variable (or also dependent variable), we used a binary classification for the values in the dataset. We employed a threshold-based labeling approach, using a three-year threshold to classify patients into two groups:  ``HF at risk" and ``not at HF risk." 
    The choice of this threshold is justified by technical reasons and clinical bases. Indeed, from a technical point of view, this threshold divides the dataset almost perfectly; in fact, we have 45\% of samples with the label ``at risk" and 55\% of samples with the label ``not at risk." This also allowed the predictive model to exhibit good prediction performance of the outcome values. From a clinical perspective, the chosen threshold aligns with a time frame during which close monitoring and proactive management are essential to mitigate adverse outcomes in patients with HF.

    
       
\end{itemize}


\subsection{Predictive Model Workflow} \label{sec:worflow}
Our predictive model (also meta-model or simply model from here onwards) is based on an ensemble learning approach, a modified stacking technique, to classify patients at HF risk or not. Standard stacking combines multiple heterogeneous models trained on the same set of features, our approach leverages domain-specific feature segregation to enhance predictive accuracy. The input data are divided into two distinct feature groups: clinical features, including demographic data, patient history, laboratory results, and risk factors, and echocardiographic features. Each feature group is processed separately using two identical ML models (\ie also clinical and echocardiographic specialized models from here onwards), each trained exclusively on one of the feature sets. This separation ensures that each model specializes in learning patterns from its respective data type without interference from other features. Both the specialized models are implemented using Logistic Regression Classifiers. In detail, we employ a five-fold cross-validation strategy using GridSearchCV, which systematically tests various combinations of hyperparameters for a split to determine the best configuration that maximizes the accuracy. In addition to these two models, we employ a third model trained on the complete set of features (\ie the merge between clinical and echocardiographic features). Indeed, we used a Random Forest Classifier to capture complex and non-linear relationships across the feature space that could be difficult to detect for Logistic regression classifiers. Using all the features together also allows for capturing interactions and patterns that span across different data types, potentially uncovering valuable insights that might be lost when the two sets of features are treated separately. A single train-test split (with 20\% of the data reserved for testing) is exploited to evaluate the performance of these three models. The selection of the three models used was informed by preliminary empirical evidence that indicated their effectiveness alone and in the combined use in the meta-model, given the specific domain of our research and the peculiarities of our dataset, . 

The outputs of the three models described above, along with their confidence in the prediction, are used as inputs for our meta-model (\ie a classifier), which is implemented as a Logistic Regression Classifier responsible for aggregating the predictions of these three models into a single prediction. The idea underlying our predictive model is inspired by the research presented by Pinem \etal~\cite{Pinem2024IntegratingMM}, where the authors trained multiple models on the same set of features to predict heart failure risk. In that work, the authors used the output of the models as input for a Voting Classifier. In contrast, our method introduces two key differences: (i)~the use of specialized models trained on disjoint subsets of features (supported by a model trained on the complete feature set) and (ii)~the use of a learnable meta-model (rather than a static voting mechanism) for the prediction.



Integrating a set of models within a meta-model offers several potential advantages over a single ML model: 

\begin{enumerate}
    \item \textbf{Interpretability and Modularity}. By separating clinical and echocardiographic features, the contribution of each data type can be analyzed independently. Additionally, a third model trained on all features captures joint relationships across the data. The study of these models separately also allows for a clearer interpretation of which type of information most influences the final predictions. We studied these aspects in our empirical assessment of the proposed solutions. 
    
    \item \textbf{Risk of Overfitting}. A single model trained on all features may overfit due to redundant or spurious correlations. By using specialized models for each group of features (clinical and echocardiographic), we reduce the risk of one type of feature dominating the learning process, improving generalization.
    
    \item \textbf{Robustness}. This structure makes the system less sensitive to missing data. If a patient lacks echocardiographic measurements, the clinical model can still contribute to predictions, preventing the complete failure of the model. The metamodel also balances predictions, reducing the impact of errors from individual models.
    
    \item \textbf{Data availability}. We built a model based on variables readily available in clinical reality.  All the components used represent part of a standard cardiac evaluation~\cite{De_Backer}.

\end{enumerate}

\section{Study Design} \label{sec:design}

In this section, we show the design of our empirical study. We took care of the suggestions/guidelines by Wohlin~\etal~\cite{WohlinBook} and Juristo and Moreno~\cite{JuristoMoreno} to design and conduct experiments in the software engineering field.

\subsection{Research Question}

The primary objective of our initial experimental assessment is to validate the meta-model for the stratification of patients at HF risk. In particular, we aimed to show whether it outperforms traditional ML models trained on the entire feature set. According to our primary objective, we defined and studied the following main Research Question:

\begin{description}
    \item[RQ] \textit{Does our predictive model outperform leading ML techniques in classifying at-risk patients?}
\end{description}

In addition to studying the three ML-based models used in our meta-model, we considered two baselines--\ie Decision Tree\footnote{It is a tree-structured model that recursively splits the data based on feature values to predict an outcome. It is intuitive and interpretable and can handle both classification and regression tasks~\cite{dt_presentation}.} and Support Vector Classifier\footnote{It is a classification variant of Support Vector Machines~\cite{svc_pres} designed to identify the optimal hyperplane that separates data points of different classes with the maximum margin. It captures non-linear relationships using kernel functions.} (SVC). We choose these baselines because they represent the standard for comparison in several AI application contexts and medical domains~\cite{ML_common,dt_example,svc_med}.


\subsection{Experimental Methodology}
To study RQ, we conducted a comparative analysis between our model and the baselines that were trained on the entire set of 33 features in Table~\ref{tab:feature_description}. All models considered in the experimental assessment were trained and evaluated using the final dataset with 20\% of the data reserved for testing. Baselines were implemented using standard algorithms from the \texttt{scikit-learn} library in Python. Just as for training the models underlying our meta-model, the baseline models have been optimized using a five-fold cross-validation strategy and GridSearchCV, which systematically tests various combinations of hyperparameters. This methodology ensures that all models were evaluated using the same data splits, enabling a fair comparison when studying RQ. 

\subsection{Metrics to Assess the Predictions of the Models }
The evaluation metrics used are standard and widely accepted in the application of ML techniques within the medical domain~\cite{Glas2003}:

\begin{itemize}
    \item \textbf{Accuracy:}  
    It represents the proportion of patients that are correctly identified, and it is computed as: $\frac{TP + TN}{TP + TN + FP + FN}\%$.
    This metric assumes values between 0\% and 100\%. The higher the value, the better it is in terms of the accuracy of the prediction. 
    For the sake of clarity and completeness, TPs (True Positives) refer to patients correctly identified as people at HF risk, while TNs (True Negatives) refer to patients correctly identified as patients not at HF risk. Instead, FPs (False Positives) represent patients mistakenly identified as at risk, while FNs (False Negatives) represent patients mistakenly identified as not at risk. 
    
    \item \textbf{Precision:}  
    It indicates the proportion of patients predicted as at risk who truly are at risk and is calculated as: $\frac{TP}{TP + FP}\%$.
    Also, precision assumes values between 0\% and 100\%. The higher the value, the better it is.
    \item \textbf{Sensitivity (or recall):}  
    It measures the proportion of actual positive patients that are correctly identified as having HF risk. It is computed as follows $\frac{TP}{TP + FN}\%$.  
    Also, this metric assumes values between 0\% and 100\%. The higher the value, the better it is.
    \item \textbf{F1-Score:}  
    It is the balanced harmonic mean of precision and sensitivity and is computed as: $2\times \frac{\text{Precision} \times \text{Sensitivity}}{\text{Precision} + \text{Sensitivity}}\%$. 
     This metric assumes values between 0\% and 100\%. The higher the value, the better it is. 
    \item \textbf{Diagnostic Odds Ratio (DOR):}  
    It is a comprehensive metric that integrates sensitivity and specificity. It is computed as follows: $\frac{TP \times TN}{FP \times FN}$. 
    The higher the value, the better it is. 
\end{itemize}

It is important to highlight that since the model is intended to stratify patients by identifying those at risk for inclusion in the telemonitoring program, minimizing FN is a primary objective. In this context, high precision is particularly desirable, as it ensures that most of the patients at risk truly need to take part in the telemonitoring program. Nevertheless, achieving high values across all evaluation metrics remains a desirable goal to ensure overall model reliability.


\begin{table}[t] 
\centering
\caption{Medical evaluation metrics of the proposed model and specialized ones}
\label{tab:metrics}
\begin{tabular}{llc}
\hline
Model                                   & \textbf{Metric}       & \textbf{Value} \\ \hline
\multirow{5}{*}{Meta-model}                 & Accuracy     & 78\%           \\
                                           & Precision    & 70\%           \\
                                           & Sensitivity  & 91\%           \\
                                           & F1-Score     & 79\%           \\
                                           & DOR                   & 20              \\ \hline
\multirow{5}{*}{Clinical model}             & Accuracy   & 72\%          \\
                                           & Precision   &  67\%     \\
                                           & Sensitivity  & 79\%       \\
                                           & F1-Score     &  78\%   \\
                                           & DOR & 16           \\ \hline
\multirow{5}{*}{Echocardiographic  model}             & Accuracy              &    78\%       \\
                                           & Precision    & 71\%            \\
                                           & Sensitivity  &  88\%       \\
                                           & F1-Score     &  78\%     \\
                                           & DOR          &  16.31          \\ \hline
\multirow{5}{*}{Random Forest Classifier}             & Accuracy              &     76\%      \\
                                           & Precision    & 69\%             \\
                                           & Sensitivity  &    88\%    \\
                                           & F1-Score     &      76\%   \\
                                           & DOR          &  14.5            \\ \hline
\end{tabular}
\end{table}

\section{Results} \label{sec:results}
In this section, we highlight the main findings. We conclude by showing possible practical implications and threats that could affect the validity of the results obtained. 

\subsection{Main Findings and Discussion}

The results of our meta-model are summarized in Table~\ref{tab:metrics} (top). Although the overall precision of 78\% appears moderate at first glance, the high sensitivity (91\%) ensures that nearly all patients at HF risk are correctly identified - an essential requirement for patient monitoring in the PrediHealth project. The precision of 70\% reflects a relatively high number of false positives, which is acceptable given the overall objective of the PrediHealth project by minimizing FNs. Furthermore, the high DOR (20) underscores the model’s strong discriminative power between patients at HF risk and those not. This reinforces the suitability of our meta-model for its deployment within the PrediHealth project. In Table~\ref{tab:metrics}, we also report the results for each model underlying our meta-model. We can note that the meta-model allows obtaining better results as compared with its underlying ML-based models. The only difference concerns the precision value of the Echocardiographic model, which is 71\%, while the precision value is 70\% for the meta-model. These results support the conclusion that combining the predictions of the three underlying models contributes to improved performance, indicating that each model captures complementary aspects of the data.



\begin{table}[t] 
\centering
\caption{Medical evaluation metrics of the baseline ML models}
\label{tab:baseline_metrics}
\begin{tabular}{llc}
\hline
ML Model                                   & \textbf{Metric}       & \textbf{Value} \\ \hline
\multirow{5}{*}{Decision Tree}             & Accuracy              & 67\%           \\
                                           & Precision   & 60\%           \\
                                           & Sensitivity  & 79\%           \\
                                           & F1-Score     & 68\%           \\
                                           & DOR                   &   4.81         \\ \hline
\multirow{5}{*}{SVC} & Accuracy              & 68\%           \\
                                           & Precision    & 63\%           \\
                                           & Sensitivity  & 72\%           \\
                                           & F1-Score    & 68\%           \\
                                           & DOR                   &    4.76           \\ \hline
\end{tabular}
\end{table}

The results in Table~\ref{tab:baseline_metrics} suggest that the meta-model outperforms the baseline models. It achieves an accuracy of 78\%, compared to 67\% for the Decision Tree and 68\% for SVC. The sensitivity of our model reaches 91\% which is well above the 79\% and 72\% for Decision Tree and SVC, respectively. This suggests a better capability of the meta-model to correctly identify patients at HF risk and not if compared with the baselines. The precision of the meta-model (70\%) is higher than the precision of the baselines (60\% and 63\% for Decision Tree and SVC, respectively).. This indicates that the meta-model has a better balance between minimizing FPs and ensuring the inclusion of at-risk patients in the telemonitoring program of PrediHealth.
Additionally, the proposed model achieves an F1-score of 79\%, outperforming Random Forest and SVC (68\% for both models). Its DOR (20) further confirms its superior ability to distinguish between patients at HF risk and those not, compared to Decision Tree (4.81) and SVC (4.76). These improvements highlight the reliability of the model for the aim for which it is developed.

\subsection{Implications}

Here, we focus on the implications of our results from the perspectives of practitioners and researchers, setting aside their specific relevance to the PrediHealth project. This choice reflects our view that the project's implications are already evident, and it allows us to emphasize that our findings hold significance beyond the boundaries of the project itself. Specifically, we developed a predictive model using a modified stacking ensemble approach applied to clinical and echocardiographic data. It achieved high sensitivity while maintaining adequate accuracy. This outcome suggests that a modified stacking ensemble approach can effectively identify a significant portion of patients at HF risk. This is relevant for {practitioners} in the healthcare field, as the model works as a valuable screening tool. Its high sensitivity supports its role in initial patient stratification. 
{Researchers} could be interested in improving our model to enhance its accuracy while maintaining high sensitivity, potentially by integrating additional data modalities or optimizing the used stacking ensemble approach. 

Our meta-model, which uses clinical and echocardiographic features and all the features flattened, presents a potential advantage, as our initial results suggest. Similar to how modularity improves software systems, this structure may enhance the interpretability of risk scores for practitioners. They could benefit from this increased transparency, as it allows them to discern whether a patient's risk is primarily driven by general clinical status or specific cardiac parameters derived from echocardiography. This aspect is also relevant for researchers. They could also be interested in empirically validating the interpretability and robustness of this architecture. For example, we provided initial evidence by showing that combining the predictions of three ML-based models by a meta-model improves the final predictions. Therefore, our empirical assessment seems to justify future research on the use of the ensemble learning approach and to explore optimal meta-learning strategies for integrating predictions from different models.

The predictive model was trained and evaluated on real patient data/records, which underwent pre-processing, including feature selection and the removal of records with missing data. Researchers could validate the performance and generalizability of the model on larger multi-center datasets representing diverse patient populations. A particular focus on HFpEF cases, as discussed in our related work, would be valuable. Researchers could also be interested in investigating the impact of different missing data imputation techniques and the development of automated, context-aware feature selection methods to enhance the robustness and reliability of a predictive model.

\subsection{Threats to Validity}
In the following section, we discuss threats that could affect the validity of our study.

\subsubsection{External validity}
Our predictive model was developed and evaluated using 357 samples. Generalizing outcomes from this clinically specific cohort poses a threat to external validity. The model's performances might differ in other populations with varying demographic profiles, healthcare practices, or different diagnoses. Furthermore, the specific set of 33 selected features might not be universally collected or hold the same predictive power in different clinical contexts. Although our study provides initial evidence for the feasibility of this model within the PrediHealth project, caution is needed when generalizing our outcomes.

\subsubsection{Internal Validity} 
The main internal validity threat of the study concerns the number of baselines used in the empirical assessment. There could be other ML models outperforming our meta-model. On the other hand, the models used as comparison baselines are among the most used in the medical field, justifying their choice. 

\subsubsection{Construct Validity}  
As for this kind of threat to validity, one key concern is how we define patients at HF risk. It may not fully capture the broader clinical view of when a patient needs intervention or faces a high risk of worsening health, which can lead to serious complications beyond death. As for the metrics used to assess our predictive model and baselines, we used standard performance measures. While these provide useful information, they possibly do not give a complete picture. Lastly, our model assumes that medical records are accurate. Any errors in these records could reduce the reliability of predictions and limit their usefulness in real-world settings.  

\subsubsection{Conclusion validity}
Our definition of the binary outcome (at risk vs. not at risk), chosen for technical reasons (dataset balance) and clinical relevance, could have affected outcomes. Using a different threshold could lead to different conclusions. Another internal validity threat concerns the choice of 33 features, which was based on clinical experience and data availability. This might leave out important features or include less useful ones.

\subsubsection{Reliability validity}
Reliability validity threats concern the possibility of replicating our study. Due to the sensitive nature of patient health information and regulations like GDPR, the raw patient-level dataset used in this study cannot be made publicly available. This inherently limits direct replication. 
We aim to make the source code implementing the data pre-processing pipeline and the described ensemble model architecture available upon reasonable request, facilitating conceptual replication and verification of our research.

\section{Conclusion} \label{sec:conclusion}

In this paper, we propose a predictive model based on machine learning (ML) techniques to identify patients at HF risk. The model adopts an ensemble learning approach using a modified stacking technique, which combines the outputs of three underlying ML-based models. We evaluate our model on a real-world dataset, and the results demonstrate strong performance in stratifying patients at HF risk. In particular, the model achieves high sensitivity (91\%), ensuring that nearly all patients are correctly identified as being at HF risk or not. The overall accuracy is 78\%, which is acceptable given our priority of minimizing false negatives (patients mistakenly identified as at risk). This is especially relevant for the telemonitoring program of the research project PrediHealth, for which the proposed predictive model has been developed. 
We also compared our model against some ML-based models, and the results indicate that our model outperforms these models.


\section*{Acknowledgment}
This project has been financially supported by the European Union NEXTGenerationEU project and by the Italian Ministry of the University and Research MUR (M4C2) as part of the project ``Tuscany Health Ecosystem'' - THE - Spoke 10 - CUP - J13C22000420001. This work was also partially supported by the project ``DHEAL – COM- Digital Health Solutions in Community Medicine" under the Innovative Health Ecosystem (PNC) - National Recovery and Resilience Plan (NRRP) program funded by the Italian Ministry of Health.

\bibliographystyle{IEEEtran}
\bibliography{bibliography}

\end{document}